\documentclass[12pt]{iopart}
\usepackage{iopams}
\usepackage{graphicx}
\begin{document}
\title[Entanglement in adiabatic cavity QED]{Entanglement in the adiabatic limit of a two-atom Tavis-Cummings model}

\author{C Lazarou and B M Garraway}

\address{Department of Physics and Astronomy, University of Sussex, Falmer, BN1 9QH, UK}
\ead{cl90@sussex.ac.uk}
%==========
% Abstract
%==========
\begin{abstract}
We study the adiabatic limit for the sequential passage of atoms through a
high-Q cavity, in the presence of frequency chirps. Despite the fact that
the adiabatic approximation might be expected to fail, we were able to show
that for proper choice of Stark-pulses this is not the case. Instead, a
connection to the resonant limit is established, where the robust creation of
entanglement is demonstrated. Recent developments in the fabrication of high-Q
cavities allow fidelities for a maximally entangled state up to $97\%$. 
\end{abstract}
\submitto{\PS}
\maketitle
%==============
% Introduction
%==============
\section{Introduction} \label{sec:1}
Entanglement, is a special type of correlation between two or
more interacting quantum systems \cite{Einstein1935,Bell1964}. 
Since they are without any classical analogy, generating
such correlations has become important for the purposes of Quantum
Computing and Quantum Information \cite{Nielsen,Galindo2002}. Among the
experimental demonstrations of entanglement, is that with atomic cavity QED 
systems in the microwave regime \cite{Raimond2001}. Apart from the generation of EPR
states \cite{Hagley1997,Osnaghi2001}, a phase gate \cite{Nogues1999,Rauschenbeutel1999}
has been realised along with the creation of Schr\"odinger cats
\cite{Brune1996}.

One of the problems encountered when considering atoms traversing a cavity
resonator, is that of the spatial effects due to the structure of the resonator
mode \cite{Yariv}. As long as the atoms are moving fast \cite{Meyer1997}, then
one can safely assume that the motion of the atomic center of mass is
classical, and utilize the spatial dependence of the coupling functions with
time dependent pulses \cite{Schlicher1989,Marr2003,Yong2007}. Based on this assumption,
in recent works we were able to study the adiabatic sequential passage of atoms
through a cavity \cite{Lazarou2008a,Lazarou2008b}. The main feature
for the system is the existence of a pure crossing in the adiabatic limit,
along with a number of possible applications in Quantum Computing, including
atomic entanglement.

Here we study the off-resonant limit for the
two-atom time-dependent Tavis-Cummings Hamiltonian \cite{Tavis1968,Tavis1969},
where now the atomic transition frequencies are subjected to time-dependent
chirps. Although one would expect the adiabatic approximation to fail, the
main result is that with a proper choice of Stark-shifting pulses, the system
follows a similar adiabatic evolution as in the resonant limit
\cite{Lazarou2008a}. In addition
to this, the use of a chirp is a robust tool for fine tuning the adiabatic
phases and consequently the output of all possible applications. 
When taking into account the recent developments in the
engineering of high quality cavities \cite{Kuhr2007,Guerlin2007}, 
we are able to demonstrate the
generation of a maximally entangled state with very high fidelity. 
 
The paper is organised as follows: In \sref{sec:2} the system and the
interaction Hamiltonian are introduced and after deriving the evolution
matrix, we demonstrate how a maximally entangled state is formed. \Sref{sec:3}
gives a brief summary of the properties of entanglement with respect to the
mode structure, and in \sref{sec:4} we discuss the importance of
spontaneous emission and decoherence. We summarize our results in \sref{sec:5}.
\begin{figure}
  \begin{center}
    \includegraphics[width=6.0cm,height=6.0cm]{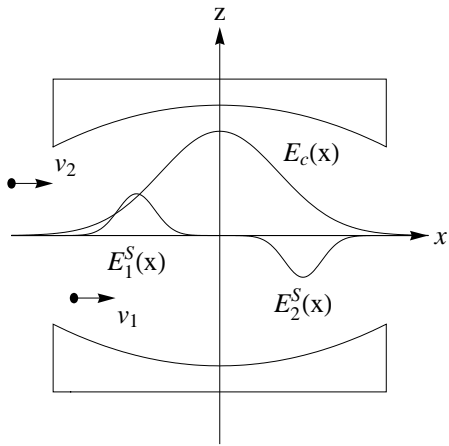}
    \caption{A pair of atoms with velocities $v_1=v_2=v$ enter a Gaussian mode
      cavity of width $2w_0$ with a delay $2\Delta t$. Two Gaussian EM pulses
      $E^S_j(x)$ are used to Stark-shift the atoms at different times. The
      $y$-axis is into the page.} \label{fig:1}
  \end{center}
\end{figure}

%================================================
% Section 2: Time-dependent Tavis-Cummings model 
%================================================   
\section{Time-dependent Tavis-Cummings model} \label{sec:2}
The system under consideration consists of two atoms sequentially crossing an
open spherical mirror cavity with velocities $v_1=v_2=v$, \fref{fig:1}. The atoms enter
the cavity with a time delay $2\Delta t$, where in addition to this they
can follow different trajectories inside the cavity. 
The Hamiltonian in the rotating wave picture and within the rotating wave
approximation reads $(\hbar=1)$ \cite{Shore}
\begin{equation} \label{eq:1}
  \hat{H}(\tau)=\sum_{j=1,2}\frac{\Delta_j(\tau)}{2}\hat{\sigma}^j_z+\sum_{j=1,2}
  \eta_j(\tau)
  \left(\hat{a}^\dagger\hat{\sigma}^j_-+\hat{a}\hat{\sigma}^j_+\right),
\end{equation}
where $\Delta_j(\tau)$ is the detuning of the $j-$th atom from the cavity mode. 

For a Gaussian mode of width $2w_0$ \cite{Yariv}, 
the coupling functions $\eta_j$ have the following form 
\begin{equation} \label{eq:3}
  \eqalign{\eta_1(\tau)=g_1\exp\left(-(\tau+\delta)^2\right),\cr\cr\eta_2(\tau)=g_2\exp
  \left(-(\tau-\delta)^2\right),}
\end{equation}
where in general $g_1\neq g_2$. The dimensionless time $\tau$ and delay 
$\delta$ are defined in terms of the interaction time $\sigma$ to be
\begin{equation} \label{eq:4}
\tau=\frac{t}{2\sigma},\qquad\delta=\frac{\Delta
  t}{2\sigma},\qquad\sigma=\frac{w_0}{v}.
\end{equation}
In deriving Hamiltonian \eref{eq:1}, we are assuming that the center of mass
motion is classical, and for this to be the case the atoms must be fast to
avoid being reflected by the mode field \cite{Meyer1997}.

For resonant interactions, i.e.\ $\Delta_j=0$, the adiabatic limit for the
above Hamiltonian proved to have a rather interesting feature
\cite{Lazarou2008a,Lazarou2008b}. At a finite
time $\tau_c=\ln(g_1/g_2)/(4\delta)$, a pure energy crossing between two of the
adiabatic states is observed. This occurs in the vicinity of a temporal 
degeneracy and results from the absence of coupling between the two degenerate
states. As a consequence of this effect, the evolution matrix has a simple
form, leading to
a conditional entanglement between the atoms and the cavity mode. Furthermore, in the
case of equal interactions, $g_1=g_2$, a number of applications in
the field of Quantum Computing such as logic gates, state mapping and
teleportation, can be realised where the output can be tuned by means of a 
single dynamical phase parameter which defines the entire system evolution.
%==================================
% Subsection 2.1: Frequency chirps
%==================================
\subsection{Frequency chirps} \label{sec:2_1}
An interesting problem that arises when considering applications based on this
system, is that of the
fine-tuning of the output state. In this paper, we show how the
Stark-shift technique, which is used to tune the interaction time 
in experiments with Rydberg atoms \cite{Raimond2001}, 
can be used to deliver a robust control over the output state.  

The whole idea requires the use of smooth Gaussian frequency chirps, such 
that each atom experiences a detuning $\Delta_j(\tau)$ at different times: 
\begin{equation} \label{eq:5}
\eqalign{\Delta_1(\tau)=\Delta_0\exp\left(-(\tau+\tau_0)^2/\sigma^2_s\right),\cr\cr
    \Delta_2(\tau)=-\Delta_0\exp\left(-(\tau-\tau_0)^2/\sigma^2_s\right).}
\end{equation}
These atomic detunings can be produced with Gaussian EM pulses. 
When the first atom enters the
cavity, a weak EM pulse with a spatial distribution $E^S_1(x)$,
\fref{fig:1}, is used to Stark-shift the atom, while crossing through the
narrow region of $E^S_1(x)$. Once the first atom has crossed that region, the EM pulse
is turned off, long before the second atom enters the cavity. In a similar
way, a second pulse
$E^S_2(x)$ is used after the first atom exits the cavity to shift the atomic
transition frequency of the second atom, \fref{fig:1}.
If both pulses have a width $2L$ and they have a peak at $x=-x_0$ and $x=x_0$
respectively, then the time $\tau_0$ and the width $\sigma_s$ will be
\begin{equation} \label{eq:6}
  \tau_0=\frac{v\Delta t+x_0}{2w_0},\qquad\sigma_s=\frac{L}{w_0}.
\end{equation}

In the adiabatic limit, the system evolution will be described by the
time-dependent eigenfunctions of the Hamiltonian \eref{eq:1}. Although the
analytic expressions for the adiabatic states can be derived, the whole process is
lengthy \cite{Lazarou2008a} and beyond the scope of this paper. 
Here we will demonstrate the
basic features of the system by means of numerical simulations with the
Schr\"odinger equation and along with simple
qualitative arguments explain the main features of the system.

Assuming that the Stark-chirps
have a short duration, $\sigma_s\ll1\quad(L\ll w_0)$, and that the location of
them, $\tau=\pm\tau_0$, is
away from the crossing point $\tau_c$ for the resonant limit, the propagator
for symmetric interactions, $g_1\approx g_2=g_0$, will have a similar form as in
the resonant limit \cite{Lazarou2008a}, i.e.\
\numparts 
\begin{eqnarray} \label{eq:7}
  &\vert n;e_1,e_2\rangle\rightarrow\vert n;e_1,e_2\rangle, \label{eq:7a} \\
  &\vert n+1;g_1,e_2\rangle\rightarrow-\vert n+1;e_1,g_2\rangle, \label{eq:7b} \\
  &\vert n+1;e_1,g_2\rangle\rightarrow\cos\left(\tilde{\phi}_n\right)\vert
  n+1;g_1,e_2\rangle-\rmi\sin\left(\tilde{\phi}_n\right)
  \vert n+2;g_1,g_2\rangle, \label{eq:7c} \\
  &\vert n+2;g_1,g_2\rangle\rightarrow-\rmi\sin\left(\tilde{\phi}_n\right)\vert
  n+1;g_1,e_2\rangle+\cos\left(\tilde{\phi}_n\right)
  \vert n+2;g_1,g_2\rangle, \label{eq:7d}
\end{eqnarray}
\endnumparts 
where in terms of the resonant dynamical phase $\phi_n$, the corresponding
off-resonant adiabatic phase $\tilde{\phi}_n$ reads
\begin{equation} \label{eq:9}
  \tilde{\phi}_n=\phi_n+2\sigma
  \int_{-\infty}^{\infty}\rmd\tau\Big(\sqrt{\Delta_1(\tau)^2+4(n+2)
    \eta_1^2(\tau)}-2\eta_1(\tau)\sqrt{n+2}\Big).
\end{equation}
The dynamical phase $\phi_{n}$, is the integral over all time for the
time-dependent energy of one of the adiabatic states in the resonant limit 
\cite{Lazarou2008a}.

The second term in \Eref{eq:9} is derived after taking into account the fact
that when $\Delta_1(\tau)$ or $\Delta_2(\tau)$ is on, then $\eta_2(\tau)=0$ or
$\eta_1(\tau)=0$ respectively. Thus the system will correspond to a single
atom interacting with the cavity mode, with a time-dependent detuning between
the atomic transition and the mode frequency. Because of
the choice made for the chirp, the adiabatic energies at this stage of the
evolution will be pushed away from each other, 
and the system will adiabatically evolve during
the time interval for which the atoms go through the fields $E^S_j(x)$.

When the chirp is off, the atom has crossed the region of the field
$E^S_j(x)$, the system returns to the initial two-atom resonant adiabatic
state, with a phase shift emerging from the chirped interaction of each atom
with the mode. Since $\Delta_1(\tau)=-\Delta_2(-\tau)$ 
and $\eta_1(\tau)=\eta_2(-\tau)$ the two contributions, one for each atom, 
can be merged into one term as in \Eref{eq:9}. Between the two chirps the
system evolves as in the resonant limit, resulting in an additional
phase $\phi_n$, first term in \Eref{eq:9}.   
 
Thus in the adiabatic limit and for symmetric interactions, the system
evolves according to \eref{eq:7a}-\eref{eq:7d}. For this to be
the case the previously derived conditions for using the adiabatic
approximation must be satisfied \cite{Lazarou2008a}. More specifically the
coupling between the atoms and the cavity mode must be strong, $g_0\gg v/w_0$,
and the delay time between the atoms must be of the order of $4w_0/v$. In
addition to these conditions , the EM fields used to produce the
Stark-chirps must be far from the cavity center, $x_0\gg w_0$, and
their width  must be smaller than the mode width, $L\ll w_0$. It is also
important to note that the pulses should be weaker than the atom-cavity 
coupling i.e.\ $\Delta_0\ll g_0$. As we see later on, violation of this
condition is not detrimental for the system.
%===========================================
% Subsection 2.2: Maximally entangled atoms
%=========================================== 
\subsection{Maximally entangled atoms} \label{sec:2_2}
\begin{figure}[!t]
  \begin{center}
    \includegraphics[width=10cm,height=6cm]{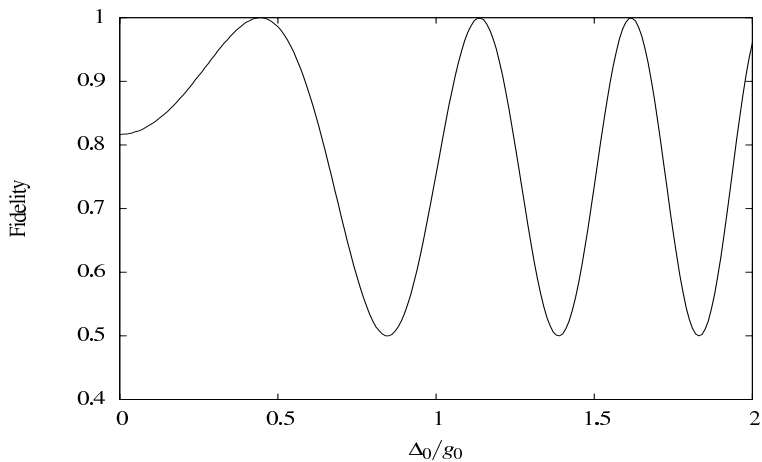}
    \caption{The fidelity for a maximally entangled state of the atoms
      \eref{eq:13} as a
      function of the chirp amplitude $\Delta_0$. The fidelity was calculated
      after numerical integration of the Schr\"odinger equation. 
      The parameters are: $g_0=30/\sigma$, $\delta=1.25$, $\tau_0=2.0$ and
      $\sigma_s=0.2$. The first peak where a maximally entangled state is
      formed is for $\Delta_0\approx0.44g_0$. The value obtained from
      Equations \eref{eq:9} and \eref{eq:14} is $\Delta_0\approx0.46g_0$.} \label{fig:2}
  \end{center}
\end{figure}
Equations \eref{eq:7a} to \eref{eq:7d} describe the conditional entanglement
of the second atom with the cavity mode. In order for this to be the case, the
atom must be initially in its ground state. Instead of entangling the cavity
mode with one of the atoms, one could generate a maximally entangled state of
the two atoms. Preparing the system in the factored state 
\begin{equation} \label{eq:12}
  \vert\psi\rangle=\frac{1}{2}\vert0\rangle\left(\vert
    g_1\rangle+\e_1\rangle\right)\left(\vert g_2\rangle+\vert
    e_2\rangle\right)
\end{equation}
before sending the atoms through the cavity, will result in a maximally
entangled state of the atoms
\begin{equation} \label{eq:13}
  \vert\psi_{en}\rangle=\frac{1}{2}\vert g_2\rangle\left(\vert
    g_1\rangle-\vert e_1\rangle\right)+\frac{1}{2}\vert
  e_2\rangle\left(\vert g_1\rangle+\vert
e_1\rangle\right),
\end{equation}
if $\tilde{\phi}_n=2m\pi$ with $m$ an integer. Thus, a maximally entangled
state of the two atoms can be generated and the output can be fine-tuned by
means of the Stark-chirps.

In \fref{fig:2} the fidelity for a maximally entangled state
\cite{Nielsen} is given for different values of the chirp amplitude
$\Delta_0$. The first important feature is the periodic reappearance of a
maximally entangled state even for $\Delta_0\geq g_0$. This is due to the fact
that the system still evolves adiabatically, even if the Stark fields are stronger
than the atom-cavity coupling. For $\Delta_0<g_0$ the phase
$\tilde{\phi}_n\propto\Delta_0^2$ and the corresponding error for the fidelity 
\begin{equation} \label{eq:14}
  F\left(\Delta_0\right)=\frac{\left\vert3+\cos\left(\tilde{\phi}_n\right)
    \right\vert}{4},
\end{equation}
will be a linear function of $\Delta_0$. Thus the Stark-shift technique is
expected to be robust for weak chirps $\Delta_0\ll g_0$. 
\begin{figure}[!t]
  \begin{center}
    \includegraphics[width=10cm,height=6cm]{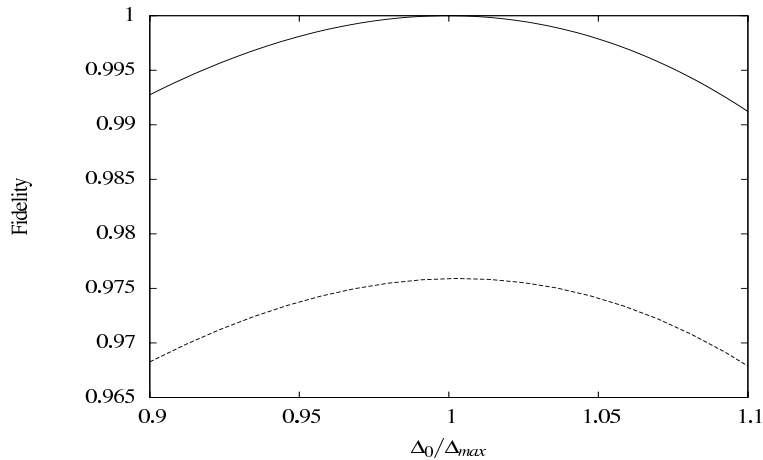}
    \caption{The fidelity for a maximally entangled state (solid)
      for variations of $\Delta_0$ around $\Delta_{max}=0.44g_0$. 
      The fidelity (dashed) for 
      a high quality cavity, $Q=4.2\times10^{10}$ \cite{Kuhr2007}, and
      circular Rydberg atoms with lifetimes $T_{at}=30\textrm{ms}$ \cite{Raimond2001}. 
      Other parameters: $g_0=30/\sigma$, $\delta=1.25$, $\tau_0=2.0$ and
      $\sigma_s=0.2$.} \label{fig:3}
  \end{center}
\end{figure}

The robustness is demonstrated in \fref{fig:3} where the
variations for the fidelity around the optimum value
$\Delta_{max}=0.44g_0$ are plotted as a function of the ratio
$\Delta_0/\Delta_{max}$. From this we can see that
variations of the order of $10\%$ around the optimum value
$\Delta_0=\Delta_{max}$ have a very small impact on the fidelity, $(<1\%)$. Thus
generating a maximally entangled state, or realising the previously proposed
applications \cite{Lazarou2008a,Lazarou2008b}, can be robust by means of weak EM
fields used to detune the atomic transition frequencies from the cavity
frequency. This in return will induce a second order shift in the dynamical
phases \eref{eq:9}, which can be used to fine-tune the output of the evolution
under the Hamiltonian \eref{eq:1} in the adiabatic limit.
%============================================
% Section 3: Entanglement spatial properties
%============================================
\section{Entanglement spatial properties} \label{sec:3}
Up to this point, one of the main assumptions was that both atoms are coupled
to the cavity field via time-dependent coupling functions with equal
amplitudes $g_1=g_2$. In the most general case and for a spherical mirror
resonator \cite{Yariv}, these amplitudes will be functions of the
coordinates $y_j$ and $z_j$, \fref{fig:1}, of the form
\begin{equation} \label{eq:15}
  g_j(y_j,z_j)=g_0\cos\left(kz_j\right)\exp\left(-y^2_j/(2w_0)^2\right),
\end{equation}
where $k=2\pi/\lambda$ is the mode wavenumber.
From this we see that $g_1\approx g_2$ if $z_1\approx z_2$ and $y_1\approx
y_2$. What is now interesting, is the properties of entanglement when this
latter condition is not satisfied.

For the resonant limit $\Delta_j=0$, asymmetries in the coupling profiles
\eref{eq:3} would introduce a second adiabatic phase $\theta_n$,
\cite{Lazarou2008b}, in addition to $\phi_n$. Despite this the conditional
entanglement between the second atom and the mode still remains, but now the output state could have two different forms each defined
by one of the two angles $\theta_n$ or $\phi_n$. This qualitative picture
holds even when the chirps \eref{eq:5} are taken into account. An important
feature, is that $\theta_n$ unlike $\phi_n$, does not 
experience any shift as $\phi_n$ does, see \Eref{eq:9}.  

It is now interesting to consider the entanglement properties for the two
atoms, as their positions $z_j$ are varied. In order to 
take into account the possibility of forming a tripartite entangled state of the
atoms with the cavity mode, we quantify entanglement in terms of the
multipartite pure state concurrence \cite{Mintert2005a}, which for our case is
\begin{equation} \label{eq:16}
  C_3\left(\vert\psi\rangle\right)=\sqrt{3-\tr\rho^2_1-\tr\rho^2_2-\tr\rho^2_c},
\end{equation}
where $\rho_j$ is the reduced density matrix for atom $j$ and $\rho_c$ is
the reduced density matrix for the cavity mode. Furthermore, we assume that 
the first atom moves
along an anti-node of the standing wave, i.e.\ $z_1=m\lambda$, while for the
second atom $z_2$ is arbitrary. With these conditions the coupling strengths will be
\begin{equation} \label{eq:17}
  g_1=g_0,\qquad g_2=g_0\cos\left(k(z_2-z_1)\right).
\end{equation}

\begin{figure}[!t]
  \begin{center}
    \includegraphics[width=10cm,height=6cm]{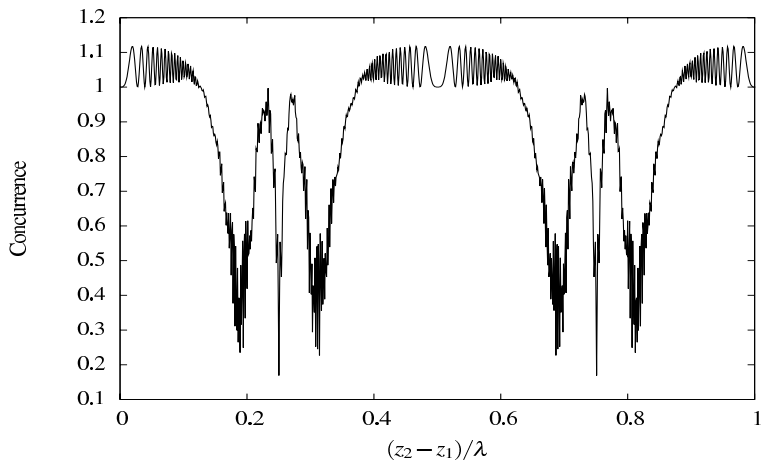}
    \caption{The concurrence \eref{eq:16} as a function of the atomic separation
      $z_2-z_1$. The initial state is $\vert\psi\rangle$
      \eref{eq:12}. The parameters are: $g_0=30/\sigma$, $\delta=1.25$,
      $\Delta_0=0.44g_0$, $\tau_0=2$ and $\sigma_s=0.2$.} \label{fig:4}
  \end{center}
\end{figure}

In \fref{fig:4}, we plot the results of a numerical simulation for the
concurrence \eref{eq:16}, as a function of the atomic separation $z_2-z_1$. As
one expects for integer or half integer values of the ratio
$(z_2-z_1)/\lambda$ we have that $g_1=\pm g_2$, and a maximally entangled state of
the two atoms is formed. These points are the local
minima where $C_3\left(\vert\psi\rangle\right)=1$. On the other hand, whenever 
the atomic separation is $(z_2-z_1)/\lambda=(2m+1)/4$, then $g_2=0$ and
this corresponds to the single atom Jaynes-Cummings model. For this limit, 
the first atom entangles with the cavity mode, forming sharp dips 
in \fref{fig:4} 
where the concurrence obtains its minimum value. Around the local minima where
a maximally entangled state of the atoms is formed, symmetric spikes with
$C_3\left(\vert\psi\rangle\right)>1$ signify the tripartite 
entanglement of the atoms and the cavity mode. For values of
$C_3\left(\vert\psi\rangle\right)<1$ and arbitrary atomic 
separations, the system could be in either a tripartite or bipartite entangled
state. In order to distinguish between the two, additional information is
necessary such as the values of $\tr\rho_j^2$ and $\tr\rho_c^2$ or the
populations and phases for the bare states. 

%===================================================
% Section 4: Spontaneous emission and cavity losses
%===================================================
\section{Spontaneous emission and cavity losses} \label{sec:4}
Spontaneous emission from the atoms to modes other than the cavity mode, and
photon losses through the cavity mirrors are both detrimental for
entanglement. In an attempt to quantify the importance of both effects, we
solved the master equation for the density matrix of the entire system,
\begin{equation} \label{eq:18}
  \frac{\rmd\rho}{\rmd
    t}=-\rmi\left[\hat{H}(t),\rho\right]+\mathcal{L}_{s}(\rho)+\mathcal{L}_c(\rho).
\end{equation}
The Liouvillian terms $\mathcal{L}_{s}(\rho)$ and $\mathcal{L}_c(\rho)$, respectively
describe the atomic spontaneous emission with a rate $\Gamma$, and
the decay of the cavity with a rate $\gamma$ into different thermal
reservoirs at zero temperature \cite{Breuer},
\begin{equation} \label{eq:19}
  \eqalign{\mathcal{L}_s(\rho)=-\frac{\Gamma}{2}\sum_{j=1,2} 
    \left(\rho\hat{\sigma}^j_+\hat{\sigma}^j_-+\hat{\sigma}^j_+\hat{\sigma}^j_-
      \rho-2\hat{\sigma}^j_-\rho\hat{\sigma}^j_+\right),\cr
    \mathcal{L}_c(\rho)=-\frac{\gamma}{2}\left(\hat{a}^\dagger
      \hat{a}\rho+\rho\hat{a}^\dagger\hat{a}-2\hat{a}\rho\hat{a}^\dagger\right).}
\end{equation}

Calculating multipartite concurrence for a mixed state is not an easy
task \cite{Carvalho2004}. One would have to perform an optimization for the
concurrence $C_3\left(\vert\psi_j\rangle\right)$ \eref{eq:16} over all pure
state ensembles $\{\vert\psi_j\rangle\}$, which equally represent the mixed
state $\rho$ that corresponds to the solution of \Eref{eq:18}. Instead of this
approach, we were able to show after a number of simulations, that the final
state of the system has a simple form
\begin{equation} \label{eq:20}
  \rho(\infty)\approx\vert0\rangle\langle0\vert\otimes\rho_a(\infty),
\end{equation} 
where $\rho_a(\infty)$ is the reduced density matrix for the atoms for
$t=\infty$. This result is for the initial state \eref{eq:12},
and is rather accurate since the average photon number $\langle
n(\infty)\rangle $ is very small, $\langle
n(\infty)\rangle<10^{-3}$, whereas the correlations between the mode and the atoms are
negligible.
This result, allows the calculation of the concurrence for the atomic pair, by
means of the two-qubit mixed state concurrence \cite{Hill1997,Wootters1998}
\begin{equation} \label{eq:21}
  c(\rho)=\textrm{max}\left\{0,\sqrt{\lambda}_1-\sqrt{\lambda}_2-\sqrt{\lambda}_3
    -\sqrt{\lambda}_4\right\},
\end{equation} 
where $\lambda_j$ are the eigenvalues of the matrix
\begin{equation} \label{eq:22}
  R(\rho)=\rho(\sigma_y\otimes\sigma_y)\rho^\ast(\sigma_y\otimes\sigma_y),
\end{equation}
where $\sigma_y$ is the Pauli matrix \cite{Merzbacher} and the eigenvalues
$\lambda_i$ are in increasing order i.e.\
$\lambda_1>\lambda_2>\lambda_3>\lambda_4$. 

\begin{figure}[!t]
  \begin{center}
    \includegraphics[width=10cm,height=6cm]{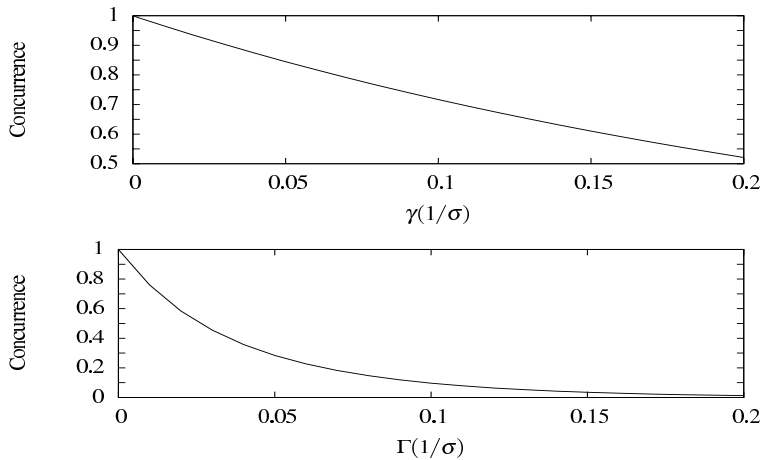}
    \caption{The two-qubit concurrence \eref{eq:21}, for different cavity decay
      rates $\gamma$ with $\Gamma=0$ (top) and for different
      spontaneous emission rates $\Gamma$ with $\gamma=0$ (bottom). Other parameters:
      $g_0\approx18.9286/\sigma$, $\delta=1.25$ and $\Delta_0=0$. The initial
      state is \eref{eq:12}, whereas for this parameters and for
      $\gamma=\Gamma=0$ the output is a maximally entangled state 
      \eref{eq:13}.} \label{fig:5}
  \end{center}
\end{figure}

The results from numerical
simulations with \Eref{eq:18} and \eref{eq:21} 
are plotted in \fref{fig:5}, where we see that
entanglement between the two atoms exponentially decays with respect to both
decay rates. Furthermore, the most important detrimental 
effect is that of spontaneous emission from the atoms. This is because the 
effective decay rate for the subsystem of the two atoms is greater
than the single atom decay rate $\Gamma$. That is, entanglement is more
sensitive to decay due to spontaneous emission than decay due to cavity losses.
%==========================================
% Subsection 4.1: Experimental feasibility
%==========================================
\subsection{Experimental feasibility} \label{sec:4_1}
For experiments with circular Rydberg states of $\textrm{Rb}^{85}$ atoms, with
lifetimes $T_{at}=30\textrm{ms}$ \cite{Raimond2001,Guerlin2007}, 
the main requirement for adiabatic evolution is for the system to be in the 
strong coupling regime, $g_0w_0/v\approx20-30$. For a coupling strength 
$g_0/2\pi=50\textrm{kHz}$ and a mode waist $2w_0=6\textrm{mm}$, 
the atomic velocities must be
$v\approx60-95\textrm{m/s}$. These speeds are large enough to avoid
reflection of the atoms from the cavity field. In addition to this the initial atomic
displacement must be $2v\Delta t\approx(4-5)w_0=(12-15)\textrm{mm}$. 

This distance
is bigger than the width of the Stark fields which is 
$2L\approx0.4w_0=1.2\textrm{mm}$. This allows the Stark fields to be
turned on and off at proper times so that they interact with only one atom and
consequently produce the desired chirps \eref{eq:5}. The location of the Stark
pulses can vary between $4.5$ and $6\textrm{mm}$ with no impact on the
robustness of the system. In addition to all these, a recent development was 
the engineering of a cavity with $Q=4.2\times10^{10}$
\cite{Kuhr2007,Guerlin2007}. For this quality factor and the corresponding
lifetimes for the circular Rydberg atoms, the fidelity for a maximally entangled
state \eref{eq:13} is reduced by about $3\%$ varying between $96$ and $97\%$,
\fref{fig:3}, which is a very high fidelity.
%============
% Conclusion
%============ 
\section{Conclusion} \label{sec:5}
In this work we have studied a time-dependent, off-resonant
two-atom Tavis-Cummings Hamiltonian. Considering pairs of slowly moving atoms,
sequentially crossing a Gaussian mode resonator, we examine the adiabatic
limit for the system. As we were able to show, for proper choice of atomic
frequency chirps the evolution of the system bears strong similarities to the
resonant limit. The system propagator has the same structure as in the
resonant limit, and one can realise the exact same applications 
\cite{Lazarou2008a,Lazarou2008b}. 

The only and also crucial difference from the resonant limit, is that the
adiabatic phase is now shifted due to frequency chirps. Because of the
linear dependence of this shift with respect
to the chirps amplitude, all the proposed applications can be robust. As an
example, the creation of a maximally entangled state of the atoms can be
achieved with fidelities higher than $99\%$. Despite potential detrimental
effects, such as atomic spontaneous emission or photon losses from the 
resonator mirrors,
recent developments in the fabrication of high quality cavities
\cite{Kuhr2007,Guerlin2007}, allow fidelities as high as $97\%$. In addition
to this, the adiabatic condition can be easily satisfied for experiments with
Rydberg atoms in the microwave regime \cite{Raimond2001}.

Although time-dependent frequency chirps could be detrimental for
the adiabatic approximation, our results show that proper timing of these chirps
could enable adiabatic evolution even if they are stronger than the
coupling between the atoms and the field. One could again 
coherently tune the output of the evolution, but this would be less
robust. This is due to the sinusoidal dependence of the error 
with respect to the chirp amplitude for strong Stark-fields. On the other
hand, for weak chirps the error in the
fidelity is very small making the Stark-shift technique highly robust. 
%============
% References 
%============    

%=========================================
%\bibliographystyle{unsrt}
%\bibliography{Papers280307.bib,Books.bib}
%=========================================
\end{document}